\documentclass[onecolumn,secnumarabic,amssymb, nobibnotes, aps, prd]{revtex4}
\usepackage{amsmath} \usepackage{amssymb}
\usepackage{graphicx} 
\usepackage{epstopdf}
\usepackage{amsmath}
\usepackage{amsmath,amssymb,amsthm,amsfonts,mathrsfs,bm,verbatim}
\usepackage{graphicx,subfigure}

\newcommand{\bea}{\begin{eqnarray}}
\newcommand{\eea}{\end{eqnarray}}

\begin{document}

\title{Thermodynamic geometry analysis of new Schwarzschild black holes}
\author{Wen-Xiang Chen$^{a}$}
\affiliation{Department of Astronomy, School of Physics and Materials Science, GuangZhou University, Guangzhou 510006, China}
\author{Yao-Guang Zheng}
\email{hesoyam12456@163.com}
\affiliation{Department of Astronomy, School of Physics and Materials Science, GuangZhou University, Guangzhou 510006, China}

\begin{abstract}
  This article puts forward a hypothesis.In this article, the derivative of the cosmological constant is positive, and there is a possibility that the constant evolves from negative in the early universe to positive in the later period. We see that when the black hole is an extreme black hole $r_{+}=r_{-}$, the curvature scalar vanishes to zero. In general, the curvature scalar is not equal to zero. When $r_{+}=3 r_{-}$, the curvature scalar is equal to negative infinity. According to Ruppeiner's theory, this time represents the phase transition of the system. It is very interesting that the divergence point of the curvature scalar happens to be the Davis transition point. We have found the phase transition point of the new Schwarzschild black hole.

\end{abstract}

\maketitle
\section{{Introduction}}
Black hole thermodynamics, or black hole mechanics, is a theory developed in the 1970s by applying the basic laws of thermodynamics to the study of black holes in the field of general relativity. Although this theory has not yet been clearly understood and formulated, the existence of black hole thermodynamics strongly suggests the deep and fundamental connection between general relativity, thermodynamics and quantum theory. Although it seems to start from the most basic principles of thermodynamics and describe the behavior of black holes under the constraints of the laws of thermodynamics through classical and semi-classical theories, its significance goes far beyond the analogy between classical thermodynamics and black holes. The nature of quantum phenomena in gravitational fields is involved.

The Ruppeiner geometry has obvious physical significance in the wave theory of thermal equilibrium. The Ruppeiner metric allows thermodynamic systems to be described by curved Riemannian manifolds, and the curvature tensor constructed by this metric has physical meaning. And it has also been shown that there is a close relationship between it and the Weinhold metric through temperature. This connection is called a conformal relation or conformal transformation:\cite{4,5,6}
\begin{equation}
d s_{R}^{2}=\frac{1}{T} d s_{W}^{2}
\end{equation}
We will discuss their conformal relationship in the discussion of RN black holes.
Since the Ruppeiner metric was proposed, many studies have been carried out on its physical significance in various thermodynamic systems. For example, classical ideal gas, ideal gas of mixed gas, ideal quantum gas, one-dimensional Ising model, van der Waals model and so on. The Ruppeiner metric was found to have a thermodynamic system phase about structural information, when the system is at phase transitions and critical points, its curvature scalar diverges (towards infinity). If the thermodynamic system has no statistical mechanical interactions (such as an ideal gas), the curvature scalar is zero and the metric is flat.

Since Ruppeiner geometry can reveal some features of statistical mechanics to some extent, we are of course interested in applying it to black hole thermodynamics. Because most of the information about a black hole is unknown to us, the model of the statistical mechanics behind its thermodynamics is not known, except for a small number of black holes in superstrings or M-theory. For the first time, Ferrara et al. introduced the method of thermodynamic geometry to study the thermodynamics of black holes. It was found that the Weinhold metric is proportional to the metric of an extreme black hole in supersymmetric mode space, which has a temperature of zero, and that the Ruppeiner metric diverges in the theory of fluctuating thermodynamic extremes. This geometric method has also been used to study BTZ black holes, RN black holes and Kerr black holes of different dimensions. They found that when the Ruppeiner metric is considered as entropy as a function of mass and charge, the metric of R N black holes is always flat and the curvature scalar equals zero. When the Ruppeiner metric is considered as a function of entropy with respect to mass and angular momentum, the metric of Kerr black holes is always curved, and the curvature scalar diverges in the case of extreme Kerr black holes. That is to say, the Ruppeiner metric of the RN black hole and the Ruppeiner metric of the Kerr black hole are different.

When $\Lambda$ is the cosmological constant about $\left(g^{\theta \theta}\right)^{2}$\cite{3}.We assume that the macro system and the micro system are closed systems, the system entropy increases to 0, the macro open system increases in entropy, and the micro system decreases in entropy. We know that Ads space can constitute Ads/CFT theory, while ds space has serious difficulties (experiments prove that the universe is ds space-time). Assuming that there is a spontaneous entropy reduction process in the microscopic system, the Ads space can evolve into a ds space. We get that (Thermodynamic geometric curvature scalar of this new Schwarzschild black hole)
\begin{equation}
\stackrel{\vee}{R}=g_{a b} R^{a b}=\frac{4 r_{+}}{\left(r_{+}-3 r_{-}\right)^{2} } .
\end{equation}
We see that although the divergence point of the curvature scalar of the Weinhold geometry is consistent with the Davis phase transition point, it cannot describe the situation of extreme black holes. And we find that the metric components of Ruppeiner geometry are not proportional to the corresponding components of the metric of Weinhold geometry, that is, Ruppeiner geometry and Weinhold geometry do not satisfy the conformal relationship.

\section{{New class of action and field equations}}
The purpose of this theory is that we find that the modified Einstein gravitational equation has a Reissner-Nodstrom solution in vacuum.
First, we can consider the following equation (modified Einstein's gravitational equation).

The proper time of spherical coordinates is\cite{1,2}
\begin{equation}
d s^{2}=G(t, r) d t^{2}-\frac{1}{G(t, r)} d r^{2}+\left[r^{2} d \theta^{2}+r^{2} \sin ^{2} \theta d \varphi^{2}\right]
\end{equation}

\begin{equation}
R_{\mu \nu}-\frac{1}{2} g_{\mu \nu} R+\Lambda (\left(g^{\theta \theta}\right)^{2})g_{\mu \nu}=-\frac{8 \pi G}{C^{4}} T_{\mu v}
\end{equation}

In this work, the action is given by the following relation  which in the special case, reduces to the Einstein-Maxwell dilaton gravity:
\begin{equation}
S=\int d^{4} x \frac{1}{16 \pi} \sqrt{-g}\left[R- \nabla_{\mu} \phi \nabla^{\mu} \phi-2 \Lambda (\left(g^{\theta \theta}\right)^{2})-e^{-2 \Phi} F_{\mu \nu} F^{\mu \nu}\right],
\end{equation}
where $\Lambda$ is a function of the Ricci scalar $R$ and $\Phi$ is the representation of the dilatonic field, also $F_{\mu \nu}=\partial_{\mu} A_{\nu}-\partial_{\nu} A_{\mu}$ (we set $8 G=c=1$ ). Variation of the action with respect to the metric $g_{\mu \nu}$, the gauge $A_{\mu}$ and dilaton field $\Phi$ gives the following field equations:
\begin{equation}
\begin{aligned}
\Lambda_{R} R_{\nu}^{\alpha} &+\left(\nabla_{\mu} \nabla^{\mu} \Lambda_{R}+\frac{1}{2} R \Lambda_{R}-\frac{1}{2} \Lambda\right) \delta_{\nu}^{\alpha}-\nabla^{\alpha} \nabla_{\nu} \Lambda_{R} \\
&=2 \nabla^{\alpha} \Phi \nabla_{\nu} \Phi+e^{-2 \Phi}\left(2 F_{\mu \lambda} F_{\nu \delta} g^{\alpha \mu} g^{\lambda \delta}-\frac{1}{2} F^{2} \delta_{\nu}^{\alpha}\right)
\end{aligned}
\end{equation}

\begin{equation}
\begin{aligned}
&\nabla_{\mu}\left(\sqrt{-g} e^{-2 \Phi} F^{\mu \nu}\right)=0 \\
&\nabla^{2} \Phi-\frac{1}{2} e^{-2 \Phi} F^{2}=0
\end{aligned}
\end{equation}
where $\Lambda_{R}=\frac{d \Lambda(R)}{d R}$ and $\nabla_{\mu} \nabla^{\mu} \Lambda_{R}=\frac{1}{\sqrt{-g}} \partial_{\mu}\left(\sqrt{-g} \partial^{\mu}\right) \Lambda_{R}$. We also have $\nabla^{\nu} \nabla_{\mu} \Lambda_{R}=$ $g^{\alpha \nu}\left[\left(\Lambda_{R}\right)_{, \mu, \alpha}-\Gamma_{\mu \alpha}^{m}\left(\Lambda_{R}\right)_{, m}\right] .$

In these equations we have:
\begin{equation}
\begin{aligned}
&\nabla_{\mu} \nabla^{\mu}\Lambda_{R}=\frac{1}{\sqrt{-g}} \partial_{r}\left(\sqrt{-g} \partial^{r}\right) \Lambda_{R}=\left(G^{\prime}\Lambda_{R}^{\prime}+G \Lambda_{R}^{\prime \prime}+\frac{G}{r} \Lambda_{R}^{\prime}\right) \\
&\nabla^{t} \nabla_{t} \Lambda_{R}=g^{t t}\left[\left(\Lambda_{R}\right)_{, t, t}-\Gamma_{t t}^{m}\left(\Lambda_{R}\right)_{, m}\right]=\frac{1}{2} G^{\prime} \Lambda_{R}^{\prime} \\
&\nabla^{r} \nabla_{r} \Lambda_{R}=g^{r r}\left[\left(\Lambda_{R}\right)_{, r, r}-\Gamma_{r r}^{m}\left(\Lambda_{R}\right)_{, m}\right]=\left(G \Lambda_{R}^{\prime \prime}+\frac{G^{\prime}}{2}\Lambda_{R}^{\prime}\right) \\
&\nabla^{\theta} \nabla_{\theta} \Lambda_{R}=g^{\theta \theta}\left[\left(\Lambda_{R}\right)_{, \theta, \theta}-\Gamma_{\theta \theta}^{m}\left(\Lambda_{R}\right)_{, m}\right]=\frac{ G}{r} \Lambda_{R}^{\prime}
\end{aligned}
\end{equation}

From the tt and rr components of the field equations, one can easily show the following relation:
\begin{equation}
\nabla^{r} \nabla_{r} \Lambda_{R}=\nabla^{t} \nabla_{t}\Lambda_{R} \Longrightarrow G \Lambda_{R}^{\prime \prime}=0 \Longrightarrow \Lambda_{R}^{\prime \prime}=0
\end{equation}
This leads to:
\begin{equation}
\Lambda_{R}=z+y r
\end{equation}
In this relation, $y$ and $z$ are just two integration constants and assumed to be positive from avoiding non-physical ambiguity. 

\section{{Thermodynamic properties of a new type of black hole}}
The new black hole is similar to the RN black hole\cite{7,8,9}, and its four-dimensional space-time metric is
\begin{equation}
d s^{2}=-G(r) d t^{2}+G(r)^{-1} d r^{2}+r^{2} d \Omega^{2},
\end{equation}
where $d \Omega^{2}$ is the line element of the unit 2D sphere, and
\begin{equation}
G(r)=1-\frac{\mu}{r}+\frac{p^{2}}{r^{2}} .
\end{equation}
where $\mu / 2=M, M$ and $p$ are the $\mathrm{ADM}$ mass and algebraic thermodynamic parameters of the black hole, respectively, here we will apply the natural unit system $\left(G=c= \hbar=k_{B}=1\right)$ . This black hole has two view surfaces, the radius of the inner view surface is $r_{-}$, and the radius of the outer view surface is $r_{+}$, and the mass and algebraic parameters are expanded into
\begin{equation}
\mu=r_{-}+r_{+}, \quad p^{2}=r_{-} r_{+}
\end{equation}
The condition $p^{2} \leq \mu^{2} / 4$ is satisfied, excluding the bare singularity case $r=0$. When $r_{-}=r_{+}$, the black hole is extreme black hole. The entropy of a black hole can be given by the area theorem
\begin{equation}
S=\frac{A}{4}=\pi r_{+}^{2} .
\end{equation}
The law of conservation of energy for a black hole is
\begin{equation}
d M=T d S+\phi d p.
\end{equation}

Therefore, we get the temperature $T$ of the black hole horizon and the algebraic parameter potential $\phi$ as
\begin{equation}
\begin{gathered}
T=\left(\frac{\partial M}{\partial S}\right)_{q}=\frac{r_{+}-r_{-}}{4 \pi r_{+}^{2} }, \\
\phi=\left(\frac{\partial M}{\partial p}\right)_{S}=\frac{\sqrt{r_{+} r_{-}}}{r_{+}}=\frac{p}{r_{+}} .
\end{gathered}
\end{equation}

We define the enthalpy $H$ of the black hole, which is expressed as
\begin{equation}
H=M-\phi p.
\end{equation}
Therefore, differentiating it to get
\begin{equation}
d H=T d S-pd \phi
\end{equation}
We can clearly establish the corresponding relationship of $(\phi, p) \rightarrow(V, P)$.
Here we define the enthalpy of the new type of black hole, including when the literature discusses Kerr black hole, RN-AdS black hole and five-dimensional black string, we also define their enthalpy, and choose the enthalpy as the thermodynamic potential.

We now begin to discuss the thermodynamic geometry of new black holes.  We first introduce the Ruppeiner metric. Based on this formula, the Ruppeiner geometric metric of the new black hole can be written as
\begin{equation}
g_{a b}^{R}=\frac{\partial^{2}}{\partial x^{a} \partial x^{b}} S(H, \phi) \quad(a, b=1 ,2)
\end{equation}
where $x^{1}=H, x^{2}=\phi$. Note that here we use the enthalpy $H$ as the thermodynamic potential, and the other parameter $\phi$ is not an extensional quantity, which is not usually used the definition of , which is a redefinition of the Ruppeiner geometric metric.

We calculate the metric as
\begin{equation}
g_{11}^{R}=\frac{8 \pi r_{+}^{2}}{\left(r_{+}-r_{-}\right)^{2}}
\end{equation}
\begin{equation}
g_{12}^{R}=\frac{16 \pi r_{+}^{2} \sqrt{r_{+} r_{-}}}{\left(r_{+}-r_{-}\right)^{2}}
\end{equation}
\begin{equation}
g_{21}^{R}=\frac{16 \pi r_{+}^{2} \sqrt{r_{+} r_{-}}}{\left(r_{+}-r_{-}\right)^{2}}
\end{equation}
\begin{equation}
g_{22}^{R}=\frac{4 \pi r_{+}^{3}\left(r_{+}+5 r_{-}\right)}{\left(r_{+}-r_{-}\right)^{2}}
\end{equation}
The curvature scalar is
\begin{equation}
\hat{R}=g_{a b} R^{a b}=-\frac{r_{+}-r_{-}}{\pi r_{+}\left(3 r_{-}-r_{+} \right)^{2}}
\end{equation}
We see that when the black hole is an extreme black hole $r_{+}=r_{-}$, the curvature scalar vanishes equal to zero. In general,
the curvature scalar is not equal to zero. When $r_{+}=3 r_{-}$, the curvature scalar is equal to negative infinity. According to Ruppeiner's theory, this time represents a phase transition of the system. It is very interesting that the divergence point of the curvature scalar happens to be the Davis transition point. We found a phase transition point for the new Schwarzschild black hole.

\section{{Summary}}
 This article puts forward a hypothesis.In this article, the derivative of the cosmological constant is positive, and there is a possibility that the constant evolves from negative in the early universe to positive in the later period. We see that when the black hole is an extreme black hole $r_{+}=r_{-}$, the curvature scalar vanishes to zero. In general, the curvature scalar is not equal to zero. When $r_{+}=3 r_{-}$, the curvature scalar is equal to negative infinity. According to Ruppeiner's theory, this time represents the phase transition of the system. It is very interesting that the divergence point of the curvature scalar happens to be the Davis transition point. We have found the phase transition point of the new Schwarzschild black hole.

\end{document}